# Quantifying Microstructure Features for High-Performance Solid Oxide Cells

C. M. Ruse[1,2,3], L. A. Hume[3], Y. Wang[1,4], T. Pesacreta[1,3], X-D. Zhou[1,4]

[1]Institute for Materials Research & Innovation, University of Louisiana at Lafayette, Lafayette, LA 70504, USA

[4]Department of Petroleum Engineering, University of Louisiana at Lafayette, Lafayette, LA 70504, USA

[3]Microscopy Center, University of Louisiana at Lafayette, Lafayette, LA 70504, USA

[4]Department of Chemical Engineering, University of Louisiana at Lafayette, Lafayette, LA 70504, USA

## 1. Introduction

The necessity to reduce fossil fuel reliance while being mindful of the growing energy demand has led to an increasing interest in the design and development of solid oxide fuel cells (SOFCs). SOFCs are highly efficient and environmentally benign electrochemical devices capable of producing electricity using various fuels (Schneider et al., 2007; Yang et al., 2013; Taylor et al., 2014). SOFCs represent an alternative to combustion processes and have twice the efficiency of internal combustion engines (> 70% in combined heart and power systems). A generic SOFC is comprised of a dense ceramic electrolyte sandwiched between two porous electrodes, whose microstructure has been studied and proved to be crucial in understanding cell performance (Vafaeenezhad et al., 2022; Virkar et al., 2000; Suzuki et al., 2009; Jiao et al., 2012, Bertei et al., 2016). Digital reconstruction gained attention in the early 2000s to better describe the reaction sites in the anode with the help of advanced Focused Ion Bean (FIB) - Scanning Electron Microscopy (SEM) (Wilson et al., 2006; Boukamp, 2006; Shearing et al., 2009; Iwai et al., 2010;

Vivet et al., 2011; Bertei et al., 2016; Jiao & Shikazono 2016) and X-ray nano-computed tomography (nano-CT) (Izzo et al., 2008; Heenan et al., 2017; Lu et al., 2017).

The microstructure of an SOFC anode, a ceramic-metal composite, is highly dependent upon the arrangement of the yttria-stabilized zirconia (YSZ) and nickel (Ni) phases (Kim et al., 1999; Stevenson et al., 2003). The functional layer of anode offers electrochemically active sites for fuel (*e.g.*, $H_2$) oxidation where the gas-filled pore space meets both the ion conductor (YSZ) and electron conductor (Ni) phases:

$$H_2(gas) + O^{2-}(YSZ) \rightarrow H_2O(gas) + 2e^-(Ni) \tag{1}$$

The junction of the three phases is known as the triple-phase boundary (TPB). A higher TPB length offers more active sites and usually is in favor of the electrochemical reaction. In addition, the size of the reaction site is determined by the volume fraction, particle size, and arrangement of the metallic and ceramic phases. Hence, accurate characterization and quantification of the TPB can help better understand the performance of electrodes with different microstructures.

Unfortunately, it has been challenging to account for actual anode microstructure by using either stereological analysis of SOFC images (Zhao et al., 2001), or theoretical 3D models (Fleig et al., 1997; Tanner et al., 1997). Wilson et al. (2006) were the first ones to demonstrate the applicability of combined ion-milling and SEM to the study of fuel cells. By combining FIB milling with SEM imaging, they were able to obtain a 3D model of the electrode microstructure with a voxel size smaller than 20 nm. They used the FIB to cut slices of a designated width through a certain layer of the cell. Then they imaged each section using SEM and used software reconstruction to produce a 3D model of the fuel cell microstructure. The authors argued that digital modeling was essential for obtaining improved estimates of the reaction site length and represented the next reasonable step in heterogenous porous media quantification analysis.

Because 3D digital reconstruction allows a microstructural comparison between different electrodes, we believed it could provide an understanding of the effect of various cell fabrication parameters on SOFC performance. The connectivity of the constituent phases of the anode is determined through the sintering process. The fabrication of SOFC typically requires a high-temperature sintering process to achieve a gas-tight electrolyte. During sintering, densification of the composite powders occurs due to grain boundary diffusion, which leads to material relocation from the surface of the particle at the grain boundary to the particle necks. Because densification is attained through particle neck growth, the particle shape is not altered (Clemmer, 2006). The sintering temperature is determined by the melting temperature of the materials used and has a direct impact on the anode microstructure (Talebi et al., 2010; Osinkin et al., 2014). An optimized sintering temperature should be high enough to densify the electrolyte layer while as low as possible to minimize anomalous grain growth. In this study, we use high-resolution FIB-SEM techniques to mill into the anode functional layer and image the exposed microstructures of two electrochemical cells prepared under different sintering temperatures. A comparison between two cells was completed to determine differences in total and connected porosity, pore and particle size, triple phase boundary (TPB) density, and hydrogen flow through the anode, based on temperature variation during sintering of the electrolyte/fuel electrode bilayers. A high electrochemical performance is rationalized with the improved microstructure by lowing the sintering temperature.

## 2.  Materials and methods

2.1. SOFC fabrication and electrochemical characterization

The SOFC substrate tape was fabricated by tape-casting process following by lamination process. The tape was cut and sintered at two different temperatures for 3 hours to obtain a multi-layered structure that consists of a $Y_{0.16}Zr_{0.84}O_{1.92}$(YSZ) electrolyte, a YSZ/NiO functional layer and a YSZ/NiO anode support layer. The YSZ/NiO substrates sintered at 1450 °C and 1365 °C are denoted as *T1* and *T2*, respectively. A $Gd_{0.2}Ce_{0.8}O_{1.9}$ interlayer was screen-printed on the YSZ electrolyte and sintered at 1200 °C. $La_{0.6}Sr_{0.4}Co_{0.2}Fe_{0.8}O_3$ was applied as the oxygen electrode with an area of 2 cm$^2$. The electrochemical performance of the cell was evaluated in the fuel cell mode with 500 sccm air and 200 sccm humidified hydrogen. The cell voltage as a function of current density was collected at a scan rate of 5mV/s by using a Biologic VMP-3 potentiostat. The electrochemical impedance spectra were acquired from 0.1 Hz to 50 kHz with an AC amplitude of 10 mA/ cm$^2$.

## 2.2. Focused Ion Beam - Scanning Electron Microscopy (FIB - SEM)

Data were obtained using a Scios 2 Dual Beam, focused ion beam (FIB), and scanning electron microscope (SEM). Both samples, *T1* and *T2*, were broken to provide a fracture face from the center of the fuel cell and one of the obtained halves was mounted with carbon tape on a 52° pre-tilted holder. A 2 nm platinum surface protection layer was deposited over the area of interest to prevent charging and a fiducial was added to aid in drift correction during milling and imaging. Trenches were created on the left and right sides of the area of interest to accommodate debris during the edge milling (**Fig. 1**).

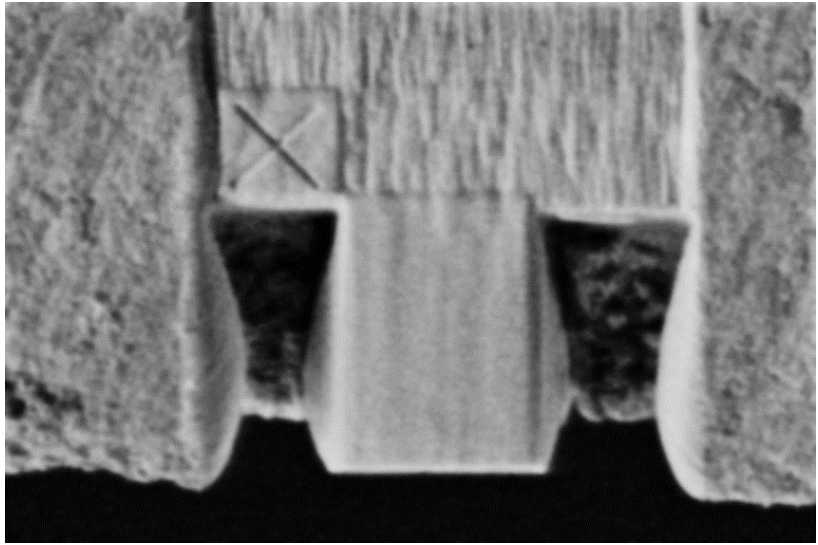

**Fig. 1.** SEM image showing the area of interest delimited by the two trenches and the fiducial used for drift corrections.

ThermoFisher Auto Slice and View software was used for milling. Focused ion beam high voltage was set to 30kV with a beam current of 5 nA. Rocking mill mode was set at a 5° tilt angle to reduce curtaining. 633 slices 20 nm apart were imaged using the backscatter electron (BSE) detector. Every second slice was captured with a resolution of 1536 x 1024 @ 8 bit and an acquisition rate of 5 microseconds, resulting in a total of 317 SEM images with a voxel size of 13.5×13.5 nm for the *T1* sample. Energy Dispersive Spectroscopy (EDS) mapping was performed on the last slice of each sample at 20 kV to determine the elemental concentrations in the phases observed in the functional layer and help recognize the three anode phases (nickel, yttrium-stabilized zirconia, and pores).

*2.3. 3D data pre-processing*

The SEM images were imported into Avizo 2021.1 to extract volumes. In the pre-processing phase, bounding areas that were highly affected by charging artifacts that could affect phase segmentation results were cropped out. This resulted in two volumes of 14.62×13.80×12.67 μm³ (for *T1*) and 18.18×11.16×12.66 μm³ (for *T2*), respectively, which were further filtered to prepare the images for the segmentation step. A Fast-Fourier Transform (FFT) stripe filter with a tolerance of 3 was used to eliminate any remaining vertical curtaining. To denoise the images, a non-local means filter with a cubic search window of 10 pixels was applied and the two volumes were resampled to cubic voxel size to ensure data compatibility with the porosity modules. The grayscale values histogram of the resampled, denoised SEM images were then used to assign a label to each pixel using a thresholding tool. Lower and upper bounds were selected on the histogram to differentiate the nickel, YSZ, and pore phases in the anode. Once phase segmentation was completed, the two samples were reconstructed in 3D to evaluate anode composition and microstructure.

### 2.4. Pore and grain network extraction

It was previously shown that anode pore and grain size distributions together with the Ni/YSZ ratio obtained after sintering has a major impact on the electrode microstructure (Holzer, 2011; Prakash, 2014). For anode pore network analysis, floating pores were removed. Connected porosity was kept by retaining all regions labeled as porosity present in two parallel planes in the desired direction. The connected porosity was then separated into individual pores to allow pore size variation analysis in the $x$-, $y$-, and $z$-directions. Pore separation served as a prerequisite for pore network modeling, which was used to approximate anode pore structure, reveal pore arrangement, pore-throat connectivity, and pore and pore throat size distribution. In addition, skeleton modeling allowed for comparison of individual pore radius. To separate the nickel and yttria-stabilized zirconia grains for statistical metrics, a watershed-based algorithm was used. The computed pore and grain radii correspond to spheres of the same volume as the analyzed objects.

*2.5. TPB identification and quantification*

To quantify the triple-phase boundary, only those locations where three anode phases met were selected. This was done by using individual segmentation labels to detect voxels that have at least one common vertex. Once an interface between connected pores, nickel, and YSZ was identified, skeletonization was performed and the TPB was displayed using spatial graph reconstruction. This allowed detailed analysis of the studied interface and quantification with respect to segment length. The total length of the electrochemical reaction site was obtained by summing up the length of all individual segments. The density was then calculated with respect to sample physical volume:

$$TPB\ density\ \left[\frac{1}{\mu m^2}\right] = \frac{TPB\ length\ [\mu m]}{Physical\ sample\ vol.[\mu m^3]} \tag{2}$$

Anode microstructure was recreated using Aviso 3D modeling, allowing us to detect the exact location of reaction sites and their extent. Hence, this technique eliminates the need for using a hypothetical microstructure where pores and grains are regarded as randomly packed spheres and subsequently, offers an accurate estimate of the triple-phase boundary by identifying its extent with respect to pore and grain distribution.

*2.6. Hydrogen flow simulation*

This is an absolute permeability experiment constrained by inlet and outlet pressures, where anode permeability is intrinsically dictated by porosity and pore size distribution. In this study, hydrogen with a viscosity of $2 \times 10^{-5}$ Pa·s was fed to the anodes of the two SOCs. Using Pergeos software, hydrogen flow through the anode was simulated in the *y*-direction, namely, from the bottom of the anode to its top. To perform the simulation, inlet pressure was considered to be 101,325 Pa and outlet pressure at the top of the anode/electrolyte boundary was set to 101,315 Pa. Darcy's law was employed to calculate anode permeability for a given gas viscosity.

## 3. Results and discussion

### 3.1. Anode phase reconstruction and quantification

The volume percentage and characteristic size of each component in the *T1* and *T2* in the *x*-, *y*-, and *z*-directions are presented in **Tables 1** and **2.** Sample *T1* has 2.5 times more yttria-stabilized zirconia (66.15%) than nickel (27.21%) with a total porosity of 7.45% although the connected porosity accounts for only 0.64% of the total volume, indicating poor pore network connectivity. The 3D representation of sample sintered at higher temperature (**Fig. 2a**) supports the values obtained and shows that the volume is dominated by the YSZ phase. In comparison, *T2* anode (**Fig. 2b**) has about equal amounts of YSZ (41.21%) and nickel (44.95%) and good porosity content (14.91%) with connected pore space accounting for 10.65% of the total volume.

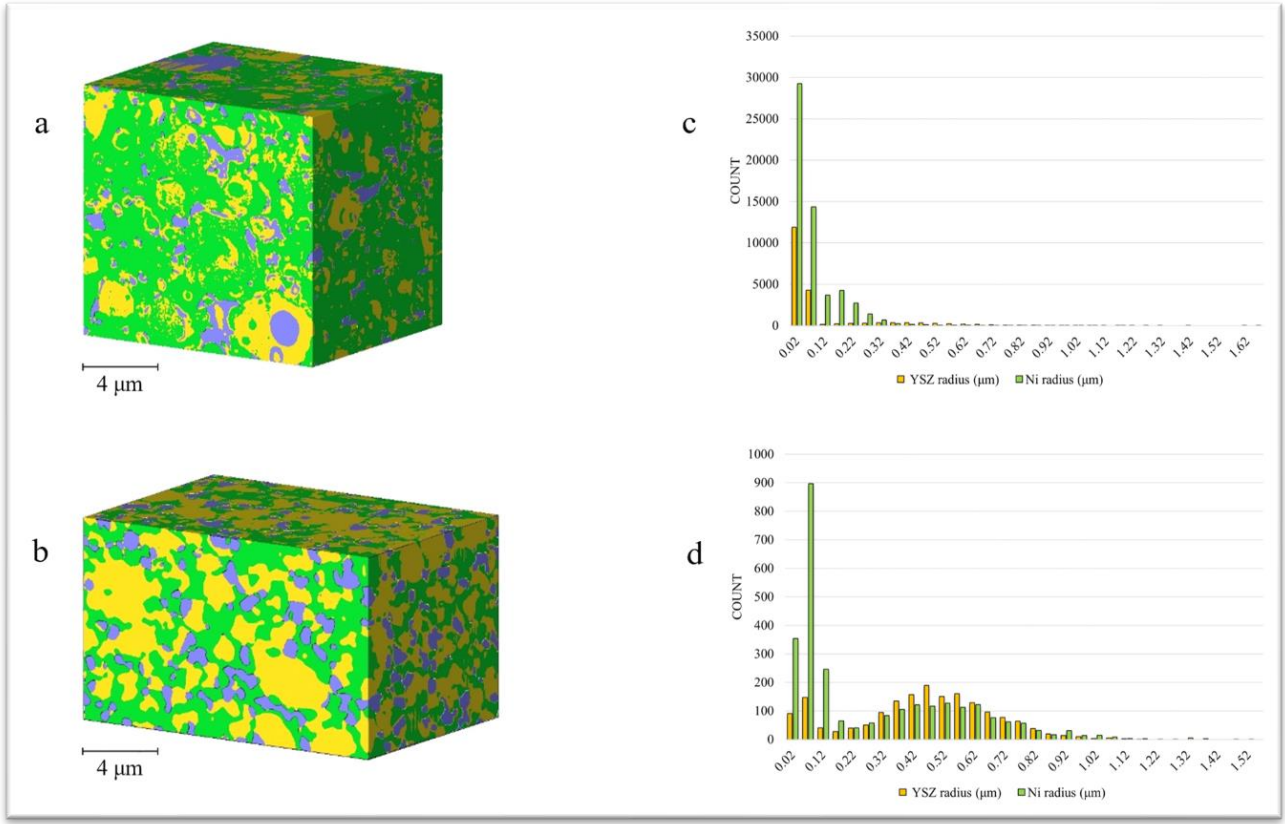

**Fig. 2.** 3D reconstruction showing the three phases of the anode of (**a**) cell *T1* and (**b**) cell *T2*. (**c**) and (**d**) show corresponding grain size distribution using histograms. YSZ is shown in green, nickel in yellow, and the pore phase in purple. The font side in FIG 2 is too small. You may plot a and b in one row and, c and d in separated rows. (More analysis on the grain size and its distribution is being conducted)

**Table 1.** Volume percent and size of the YSZ, nickel, and total and connected porosity phases in sample *T1*. The size of the pores was quantified using the connected pore network and thus, no values are shown for the total porosity.

| | | Volume percent (%) | | | | Size (μm) | |
|---|---|---|---|---|---|---|---|
| *T1* | *Anode phase* | **Mean** | **STD** *x*-direction | **STD** *y*-direction | **STD** *z*-direction | **Mean** | **STD** |

| | | | | | | |
|---|---|---|---|---|---|---|
| YSZ | 66.15 | 3.60 | 4.20 | 4.36 | 0.08 | 0.17 |
| Ni | 27.21 | 3.86 | 4.10 | 4.12 | 0.06 | 0.08 |
| Total porosity | 7.45 | 1.13 | 1.26 | 1.08 | - | - |
| Connected porosity | 0.64 | 0.85 | 0.66 | 0.37 | 0.14 | 0.08 |

**Table 2.** Volume percent mean and variation in the *x*-, *y*-, and *z*-directions along with particle size in sample *T2*. The connected pores were used to quantify pore size.

| | | Volume percent (%) | | | | Size (μm) | |
|---|---|---|---|---|---|---|---|
| | *Anode phase* | Mean | STD *x*-direction | STD *y*-direction | STD *z*-direction | Mean | STD |
| *T2* | YSZ | 41.21 | 3.53 | 2.01 | 4.92 | 0.42 | 0.23 |
| | Ni | 44.95 | 3.43 | 2.68 | 5.61 | 0.26 | 0.28 |
| | Total porosity | 14.91 | 1.56 | 1.40 | 1.55 | - | - |
| | Connected porosity | 10.65 | 3.08 | 1.36 | 2.64 | 0.31 | 0.11 |

Volume segmentation revealed that nickel particles had clear boundaries and were easily distinguishable, but the ceramic phase was closely compacted and comprised of fused grains. The metallic phase of the anode was porous and easily observed (**Fig. 2b**). **Tables 1** and **2** show that the YSZ grains were larger than the nickel particles in both samples, which agrees with observations made by Chen et al. (2007). The authors explained that smaller nickel grains are better because they can contribute to mitigating the thermal expansion mismatch. Grain size distributions in **Figs. 2c** and **2d** appear to be skewed, which indicates that the standard deviation cannot be treated as a measure of data spread in both directions. Sample *T1* was characterized by grain radius distributions with long right tails, which is caused by a population of relatively large grains. In this case, the standard deviation can be used to assess the spread of the data in the right direction. While the mean of the YSZ grain size does not exceed 0.08 μm in sample *T1*, the anode of cell *T2* is characterized by metallic and ceramic phases with larger particles. The yttria-stabilized zirconia has a mean grain radius of 0.42 μm. More grain size and its distribution is being conducted and results will be updated.

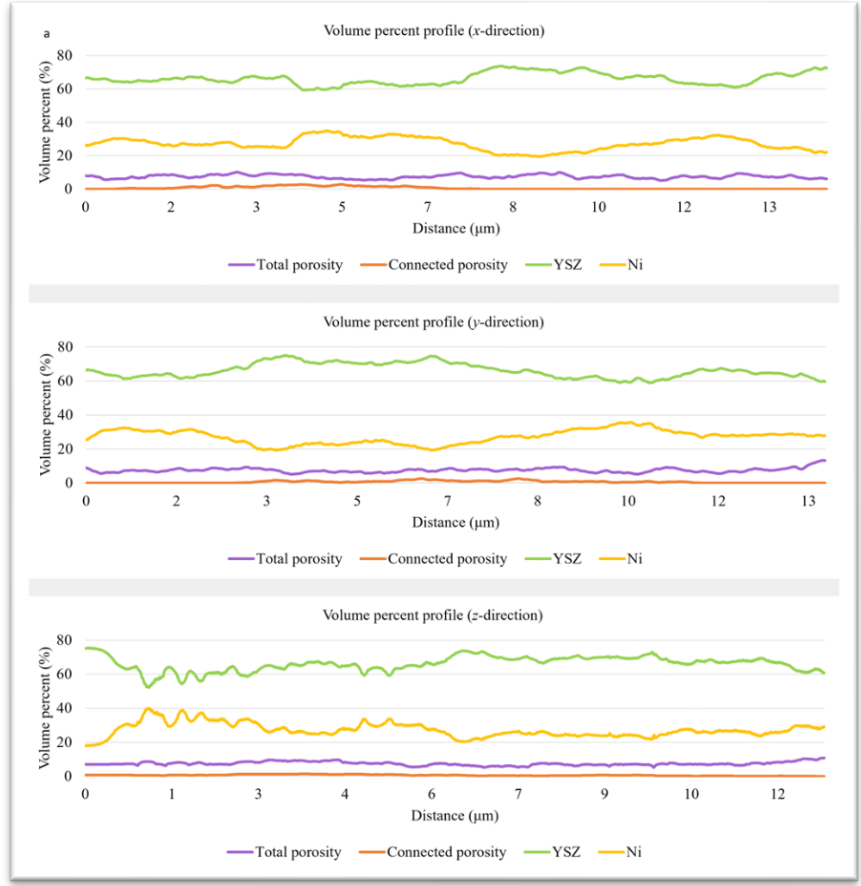

**Fig. 3.** From top to bottom: anode phase volume percent variation profiles in the *x*-, *y*-, and *z*-directions obtained for cell *T1*. Total porosity is shown in purple, connected porosity in orange, YSZ in green, and nickel in yellow.

Phase variation across the two volumes was studied in **Figs. 3** and **4** and provides insights into the dependencies between the three phases. The three profiles can be interpreted in the following manner: the profile in the *x*-direction shows volume percent variation parallel to the anode/electrolyte boundary, while the second profile presents phase volume changes from the inlet to the outlet of the anode. To study compositional changes across the thickness of the anode, a third profile in the $z$-direction of the volume is created. The two profiles in the $y$-direction show a moderate correlation between the nickel and porosity volumes, with pore space increasing where less nickel is present. This is a direct consequence of nickel particles being replaced by pore space. A correlation coefficient of -0.51 describes the relationship between the nickel and connected porosity volumes in sample *T1*. Total porosity shows a slightly stronger negative correlation to the nickel in sample *T2*, with a correlation coefficient of -0.61. While nickel and porosity volumes are inversely proportional along all directions in sample *T2*, the two phases can be directly proportional in sample *T1*. According to the statistical analysis (**Tables 1** and **2**), both the ceramic and metallic phases show greater variation across the anode thickness, while connected porosity varies more in the direction parallel to the anode/electrolyte boundary. The connected porous space is characterized by standard deviations of 0.85 and 3.08 in samples *T1* and *T2*, respectively.

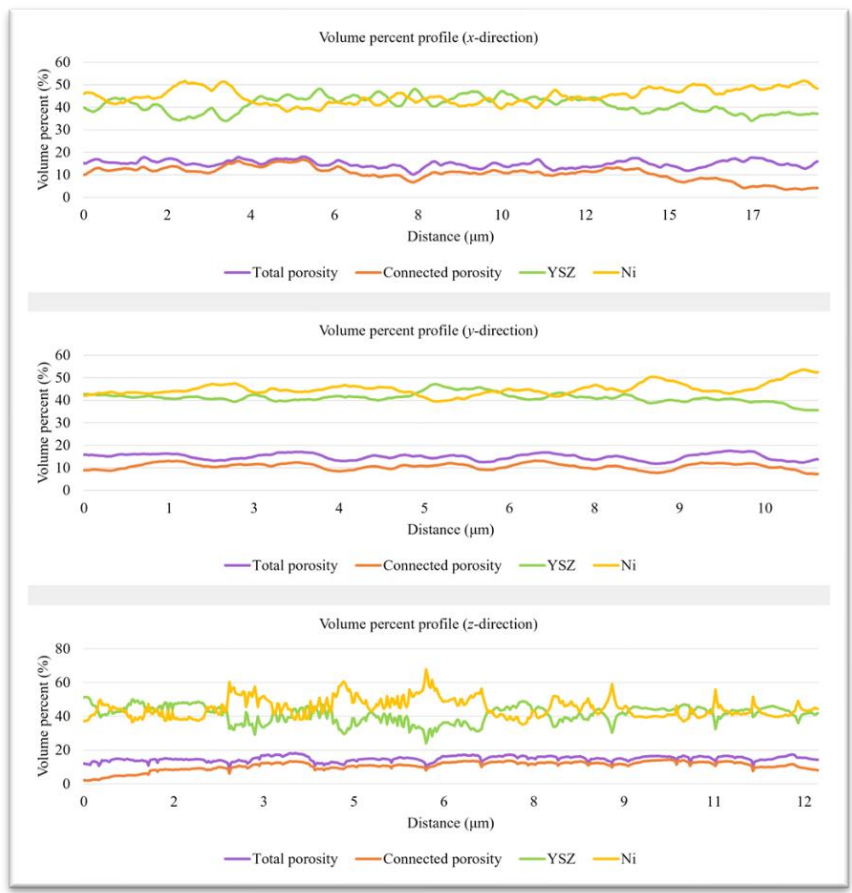

**Fig. 4.** From top to bottom: anode phase volume percent variation profiles in the *x*-, *y*-, and *z*-directions obtained for cell *T2*. Total porosity is shown in purple, connected porosity in orange, YSZ in green, and nickel in yellow.

For a comparative, 2-D analysis, samples were obtained from 5 different regions from each sample to investigate porosity volume. A mean of 13.18% with a standard deviation of 1.27% was obtained for *T1*, and a mean of 9.20% and a standard deviation of 1.55% for *T2*. The standard deviations of the studied volumes have values similar to the ones obtained for the 3-D analysis of *T1* and *T2* (**Tables 1** and **2**). This appears to be a relatively quick and accurate way to obtain limited data without the time and expense of milling.

*3.2. Connected pore space modeling for TPB characterization*

One prerequisite of an active electrochemical reaction site is for the porous space to be connected to allow uninterrupted hydrogen flow from the inlet to the outlet of the anode. For *T1* and *T2* the connected pore space was separated from the initial total porosity and the obtained structure was studied. In the *T1* sample across the *y*-direction, connected pore space fails to link the inlet to the outlet (**Fig. 6a**). But, sample *T2* is characterized by the presence of a robust connected pore network in all directions (**Fig. 6b**). To reconstruct the connected pore space of *T1* and *T2*, a region of interest (ROI) delineated by the limits of the connected pores was selected. The obtained pore network model (**Fig. 5**) shows that the *T1* network has pore with radii between 0.02 and 0.46 μm with a mean of 0.14 μm and the pore throats are almost two times smaller than the pores. The *T1* distributions are displayed in **Fig. 5c,** and similarly to the grain size distributions, are right-skewed. The *T2* pore network pertaining to the sample sintered at a lower temperature is characterized by normally distributed pore space with bigger pores of up to 1.52 μm in diameter and a mean of 0.31 μm. But, interestingly, as is the case for *T1*, the throats of the *T2* network are two times smaller than the pores with a mean throat size of 0.15 μm.

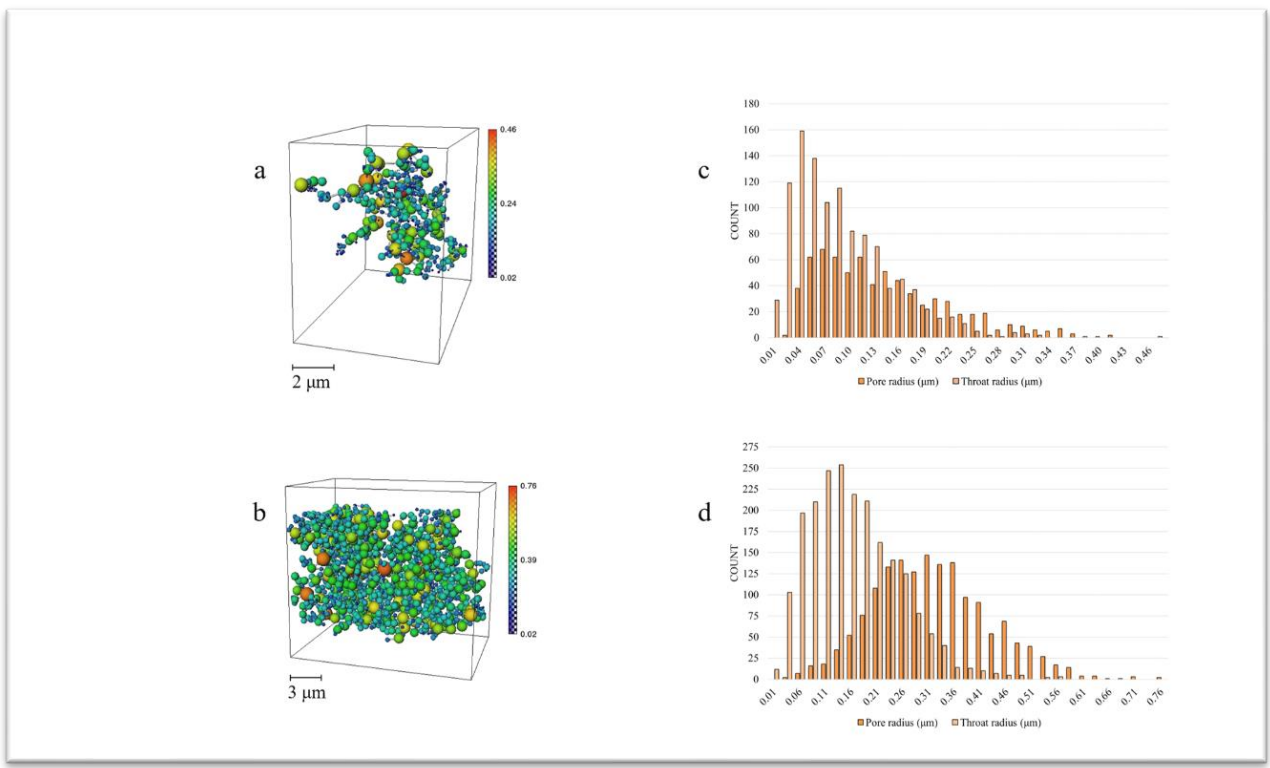

**Fig. 5.** Connected pore network representation using estimated pore size for (**a**) cell *T1* and (**b**) cell *T2*. Pore radius scales are displayed on the right side of the images. (**c**) and (**d**) show pore size and pore throat size distribution using histograms for *T1* and *T2* respectively. Orange is used for the pores and light orange for the throats of the network.

The triple phase boundary comprises all active reaction sites where the connected pore space meets with the metallic and ceramic phases of the anode. The connected porosity only was used to identify such sites and represent their length using individual segments. The TPB is displayed in **Figs. 6c** and **6d**, and encompasses all triple junctions where the boundaries of the three phases meet. The lengths of the separate segments were summed up to obtain the total length of the reaction site in each sample. For the ROI corresponding to cell *T1*, the TPB has a length of 1559.62 μm which results in a density of 0.61 μm$^{-2}$ for a total sample volume of 2556.04 μm$^3$. In contrast, sample *T2* is characterized by a much higher TPB density of 3.92 μm$^{-2}$ corresponding to a sample volume of 2568.59 μm³. The higher TPB density of the second sample is the result of a more extensive pore network and a balanced Ni/YSZ volume ratio. The increased sintering temperature used for T1 apparently led to the formation of very small nickel particles unable to support porosity formation, which subsequently led to a significant decrease in the electrochemical reaction site.

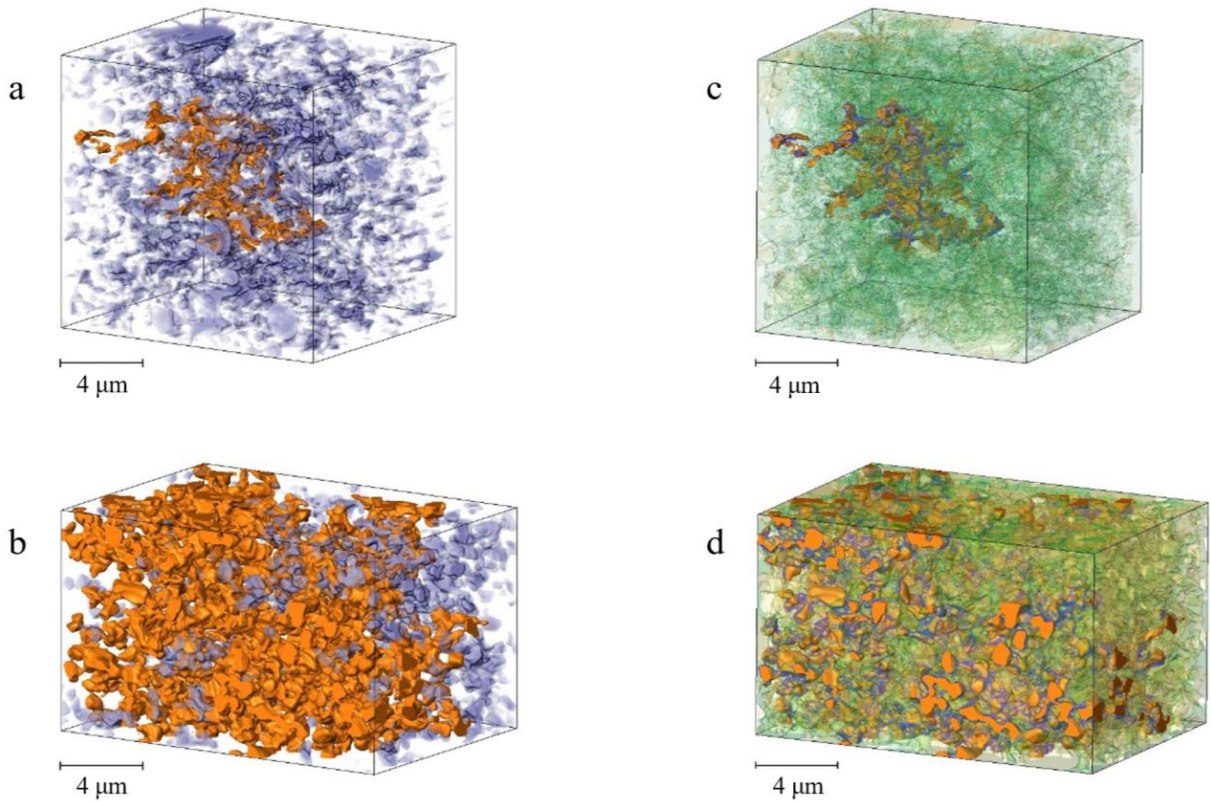

**Fig. 6.** 3D reconstruction of the connected pores (in orange) and floating pore space (in purple) for (**a**) cell *T1* and (**b**) cell *T2*. The corresponding triple-phase boundaries are displayed on top of the connected pore space in images (**c**) and (**d**), respectively.

**Table 3**. Triple-phase boundary parameters for the studied cells.

|  | *T1* | *T2* |
|---|---|---|
| *Sample volume (μm³)* | 2556.04 | 2568.59 |
| *TPB length (µm)* | 1559.62 | 10062.40 |
| *TPB density (µm⁻²)* | 0.61 | 3.92 |

*3.3. Anode permeability and cell performance*

Hydrogen flow simulation was performed for regions corresponding to connected pore space. The anode phase arrangement in sample *T1* led to a drastic reduction in the region of interest due to the poorly connected porosity network. **Fig. 7a** shows that flow is restricted to a very small area which is in good agreement with the very low permeability obtained for this anode volume ($k = 9.58 \times 10^{-5}$ md). Alternatively, sample *T2* exhibits well-developed flow within the entire studied volume with an anode permeability of $k = 8 \times 10^{-3}$ md. **Fig. 7b** presents the pressure field evolution from the inlet (higher pressure) to the outlet (lower pressure) within the investigated anode region. Electrode permeability can be further used to explain the performance of the two cells.

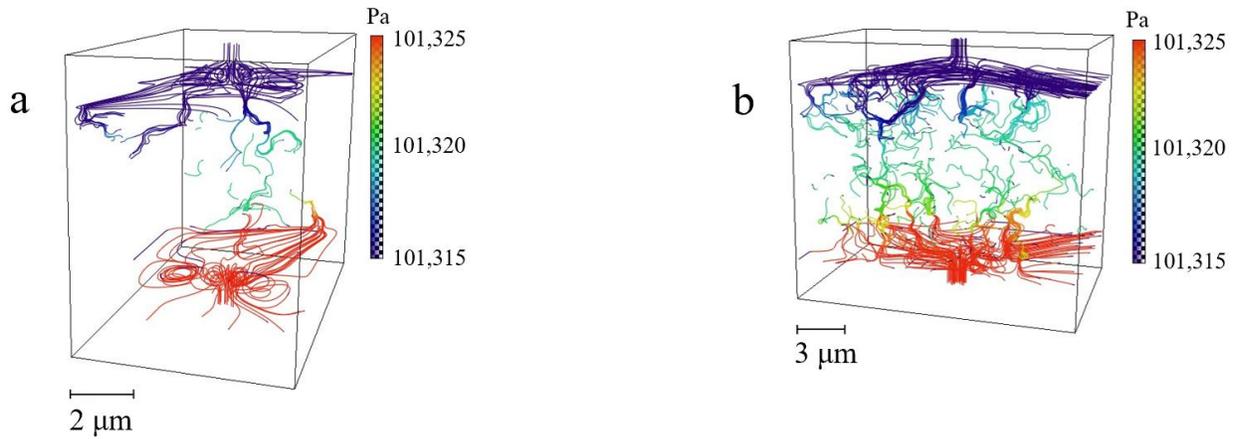

**Fig. 7.** Pressure field evolution inside the connected network of the two anode samples. (**a**) Cell *T1* shows restricted hydrogen flow, while (**b**) *T2* exhibits good pore connectivity which allows gas flow from the inlet to the outlet of the studied volume.

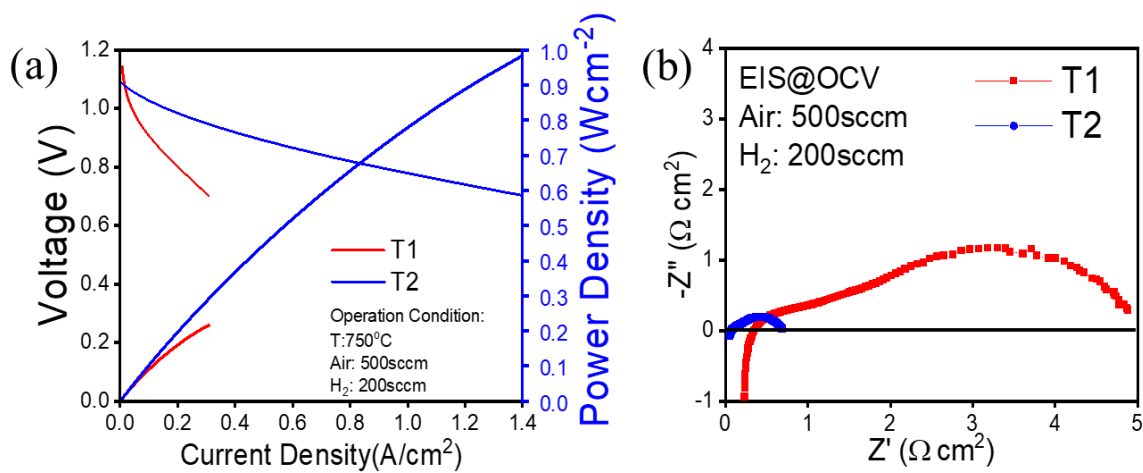

**Fig. 8.** (**a**) Current density-voltage-power density relation of two SOFCs (*T1* and *T2*) sintered at different temperatures (**b**) Electrochemical impedance spectra (EIS) at open circuit of two SOFCs (*T1* and *T2*) sintered at different temperatures.

Fig 8 (a) shows current density-voltage-power density (i-V-P) relation of two cells. Cell *T1* is characterized by low power density at 0.75V (0.19 W/cm$^2$) while cell *T2* has a power density of 0.86 W/cm$^2$. Both cells have high open circuit voltages (>1.09V), which indicates a densified electrolyte and a good seal. Electrochemical impedance spectra (EIS) (Fig 8 (b)) shows that Cell *T1* has higher total cell resistance (4.88 $\Omega \cdot$ cm$^2$) than Cell *T2* (0.69 $\Omega \cdot$ cm$^2$), which is generally contributed by the polarization resistance of the electrodes. These characteristics are a result of the anode microstructure obtained using different sintering temperatures. Phase arrangement in sample *T1* was heavily impacted by the relatively high temperature used during sintering, leading to small pores in a compact mass of yttria-stabilized zirconia with little nickel. Lack of good connectivity between the pores impacts the permeability of the network and hinders hydrogen flow in the *T1* cell and results in a high diffusion impedance shown as the low frequency arch. This agrees with Geagea et al. (2015), who explained that high permeabilities are associated with robust gas paths capable of reducing diffusional polarization loss. In addition, the low TPB length results in a slow kinetics in the whole functionally layer, which further increases the polarization impedance. The good performance of cell *T2* is confirmed by the presence of a robust pore network and extended TPB length, as revealed by the FIB-SEM reconstruction.

## 4. Conclusions

Focused ion beam (FIB) – scanning electron microscopy (SEM) allowed the characterization of the microstructure of two solid oxide fuel cells prepared at different sintering temperatures. 3D volume reconstruction showed that a relatively low sintering temperature significantly and positively affected  distribution, volume and particle size of yttria-stabilized zirconia, nickel, and pore phases inside the anode, as well as the extent of the important triple-phase boundary interface. The poor performance of the *T1* sample sintered at a higher temperature is explained by the poorly connected pore network and very low-density triple-phase boundary. The pore space inside the *T1* anode was unable to ensure continuous hydrogen flow from the inlet to the outlet and thus exhibited very low gas permeability. In contrast, the *T2* sample sintered at a lower temperature had approximately equal amounts of YSZ and nickel and larger pores, which allowed formation of significantly more TPB electrochemical reaction sites. The higher power density of the *T2* cell was also the result of its robust pore network capable of transporting hydrogen throughout the anode. The methodology used in this paper eliminates the need for employing hypothetical structures and provides accurate estimates of the investigated parameters by evaluating microstructures that were successfully reconstructed using high-resolution microscopy techniques.

## Acknowledgement

We would like to thank the U.S. National Science Foundation for the support of this work under NSF-1920166 and U.S. Department of Energy for the support of this work by the Office of Fossil Energy and Carbon Management under DE-FE0032110.